# Taxonomy of Digital Forensics: Investigation Tools and Challenges

Nikita Rana[1], Gunjan Sansanwal[1], Kiran Khatter[1,2] and Sukhdev Singh[1,2]
[1]Department of Computer Science and Engineering
Manav Rachna International University, Faridabad-121004, India
[2]Accendere Knowledge Management Services Pvt. Ltd., India

**Abstract**

In today's world of computers, any kind of information can be made available within few clicks for different endeavours. The information may be tampered by changing the statistical properties and can be further used for criminal activities. These days, Cybercrimes are happening at a very large scale, and possess big threats to the security of an individual, firm, industry and even to developed countries. To combat such crimes, law enforcement agencies and investment institutions are incorporating supportive examination policies, procedures and protocols to address the complete investigation process. The paper entails a detailed review of several cybercrimes followed by various digital forensics processes involved in the cybercrime investigation. Further various digital forensics tools with detail explanation are discussed with their advantages, disadvantages, challenges, and drawbacks. A comparison among all the selected tools is also presented. Finally the paper recommends the need of training programs for the first responder and judgement of signature based image authentication.

## 1. INTRODUCTION

Computer is a masterpiece made by the human race that has made our lives smooth and effortless. Computers have become the very bedrock of today's technological environment and we use them in almost every aspect of our customary life. They are everywhere from shopping, banking to school and hospitals, even our own homes. Businesses depend on these devices and the Internet to do their daily transactions, marketing and communications across the globe and given to our desideratum to have the best of everything it has seen noteworthy diversifications. Generous volume of information that includes financial and personal information is stacked on the servers daily. People have become so reliant over the usage of computers that they have started to call this structured form of environment as Cyberspace. The word 'Cyberspace' itself defines the space of cyber or in technical terms, a notional environment of computer networks over which communication takes place [1]. The term was first featured in a book called "Neuromancer" by William Gibson in 1984 (Page 4, Phantasia Press Edition, Bloomfield, MI, 1986) [2]. It is a virtual computer world which is used to establish worldwide online communication [3]. It engages TCP/IP protocol to facilitate data exchange exercises. Although it seems like an abstract point, cyberspace has become the destination of millions of people who visit it every day to share their ideas and thoughts, play games, socialise, initiate and boost business, and, many other activities, for this it looks like it must be a real place [4]. Just like the way, every real place is valued in the market; cyberspace has also gained its own value, which can be seen in the form of various applications and services people use in their everyday lives. If we speak in terms of limitations, it has no physical barriers and hence, is not restricted by any territorial limits. Today, a Service provider in one part of the world can freely offer his services to a person sitting across the globe. It is a levelling medium where each person connected to the Internet can avail everything that the Internet has to offer and reap its benefits. The everyday scenarios can be that of people no longer having to opt for hitchhiking and can get the benefits from the application driven cab services such as Ola and Uber. These start-ups have made the drivers to become entrepreneurs without having to become a part of taxi services. Additionally, people can get free knowledge and information from the materials available over the cyberspace. The advent of online education platforms such as Quora, Udemy, Coursera etc. have enabled people to access knowledge which was earlier localized in universities and schools. Even the government, has attained various benefits from the advent of cyberspace as all the form submissions such as voter id cars, driving license, telephone bill and other payments can be done over cyberspace without the people having to wait in long queues. Cyberspace has issued a sense of transparency and the removal of middlemen between the citizens and the government. However, even if the cyberspace has so many benefits it cannot be considered fully trustworthy. Just like we say a coin has two faces, cyberspace also has two faces, one where there are thousands of people getting befitted from it and the other where lots of people who are getting harmed because of it. With the increase in the scope of computer networks, the world has seen a rise in computer oriented crimes alternatively referred as cyber crimes **[5]**. 56% of the participants in a survey conducted by PWC observed an increase in the level of cyber crime **[6]**. This leads one to conclude that cybercrime never came days or years after the invention of cyberspace rather it came with it. The real challenge with cybercrime is that the accused or the criminal can stay hidden in the virtual domain **[7].**

   Cyber crimes also known as e-crimes, hi-tech crimes or electronic crimes is an operation performed by an individual which have some or intensive knowledge of the computer and its whole system with the aim of corrupting, illegally extracting the data, deleting the stored data. If the word 'cybercrime' is searched upon the Internet we will come across various definitions of it. The Cyber.laws website defines cyber crimes as "a crimes or any illegal activity that involves a network and a coordinating computer" **[8]** whereas Norton by Symantec defines it as "a crime that has some kind of computer or cyber aspect to it" **[9]**. Whatever be the types of definitions provided by various sources one factor remains the same that cybercrime unlike traditional crimes, provides a major hindrance in unveiling the criminals as the user's identity may be hidden or fraud over the virtual domain **[10]**.
These crimes happening over the cyberspace have impacted huge masses to such an extent that billions of dollars have been nagged by the criminals from the organizations without them knowing until an extreme damage has been done to their part. Therefore, as the extent of damage increases the need to solve such an extent also rises, which is where we come across something which can be related to the long heard tales of Sherlock Holmes by Sir Arthur Conan Doyle **[11]**. The series' character has been imprinted in the minds of millions along with the fact that forensic sciences along with logical reasoning can be considered as the primary foundation for solving many cases, has made many dream about becoming detectives someday. Just like the way forensic science is used in a number of traditional cases it can also be used in the crimes which happen over computers or cyberspace. Such an investigative method is known as computer or digital forensics.

   Prior moving ahead in explaining the details of digital forensics and its role in combating cybercrimes, it's imperative to understand the types of cybercrimes and the huge gap between the occurrence of cybercrimes and the ability of the legislation to solve them.

# TYPES OF CYBERCRIME

| AUTHOR | TYPE OF CYBERCRIME. | EXAMPLES |
|---|---|---|
| Prosise, Chris, Kevin Mandia, and Matt Pepe [27] | **Unauthorized access:** Accessing or intercepting any piece of data or information that one is not supposed to see or is unauthorized to review [27,28]. | A man named kumar gained unauthorized access to Joint Academic Network(JANET) and altered the contents of the files, changed passwords, deleted and added few files so that access can be denied to the authorized user. In order to do he was illegally accessing the BSNL broadband connection and modifying the computer database [12]. |
| Margaret Rouse[30] | **Identity theft:** When an individual acquire or impersonates someone else's identity. In other words, pretending to be someone else [29,30,33]. | • In September of 2010 a complaint was registered by the President household and Secretariat against an imposter who created a fake profile in the name of the Indian president Pratibha Patil on facebook [15,16]. <br><br>• In 2011, a fake twitter account of Sarah Palin, the former governor of Alaska was created, which tweeted out open invites to the governor's house for a barbecue. This tweet led to a situation which was difficult to handle by the governor's security [20]. |
| Vangie Beal [31] | **Phishing:** Targeting individuals or group of individuals and luring them into giving up their personal and/or financial information and using the acquired information for own benefit. Phishing can be of different types namely clone phishing, spear phishing etc [31]. | • The customers of ICICI bank were sent emails in which they were asked to change their username and passwords. The emails resembled the original mails from the bank so the customers clicked on the url mentioned in the mail thus giving up their bank details for the attackers to exploit [12, 21]. <br><br>• LinkedIn website was cloned in order to steal the credentials of the users. This was the "Ham-Fisted" phishing attack [23]. <br><br>• In December, 2013 a man was accused and arrested for initiating a phishing attack which targeted college students in UK . The students were asked to update their loan details on a fake website that resulted in the loss of money from the student's accounts [22]. |
| Mindi McDowell [35] | **Denial of service:** Overburdening a system by sending too many requests at once which result in failure to complete normal requests. When too many requests come at once the system is unable to recognise the priority of the task and thus is unable to do the normal functioning and focus on the extra information coming in [34, 35]. | • A foreign citizen residing in Shimla for past thirty years wanted to buy land in shimla which was offered by the shimla housing board at very low rates ,but his application was rejected because he was not an Indian citizen. To take revenge he sent thousands of email to the shimla housing board's address and this ultimately lead to the crash of their servers [12]. <br><br>• Boston children's hospital was targeted by a hacker group in 2014. It became the first healthcare organization to encounter such |

| | | attack. The attackers targeted the hospital using the distributed denial of service attack. The hospital shared the same internet service provider (ISP) with seven other healthcare organizations. This attack was intended to bring down multiple healthcare organizations in Boston **[24]**. |
|---|---|---|
| Ricardo A. Morales<br><br>Sarita Hill Coletrane<br><br>Harold Shreves<br><br>Jeffrey Butterworth<br><br>Aaron M. Stutz **[38]** | **Cyber bullying or cyber stalking:** Tormenting or stalking someone online. It means that a person may harass someone for personal vengeance or for a mere purpose of poking fun **[28]**. | • A student studying law in Delhi University was accused and punished for stalking and bullying a woman online after she refused to his proposal. He created fake profile of the lady on social site to defame her **[17]**.<br><br>• In 2010, an 18 year old boy Tyler Clementi who was a student at Rutgers University committed suicide by jumping from the George Washington bridge after being electronically stalked and bullied by his roommate Dharun Ravi. Ravi spied on Tyler using a webcam and recorded an intimate encounter with the same-sex and posted the video online. Tyler was bullied because of the video and took his own life **[25,26]**. |
| Cristina Chipurici **[41]** | **Fraud** Manipulating or moulding financial data in order to benefit a particular individual or organisation. By acquiring the financial records the attacker may use the money or the statement details **[39, 40, 41]**. | • Five people who were working in MsourcE in pune were arrested in April 2005 for alleged fraud of 425,000 dollars from the account of four New York based holders which had their account in Citibank **[12, 18]**. |
| Subhojyoti Acharya **[36]** | **Cyber terrorism** Blackmailing or attacking a selected organisation say the government into following some demand and brainwashing the younger generation to follow the path of terrorism via any online medium. The terrorist groups use social media websites as a platform to spread terrorism and anti-government ideology **[36]**. | • In May of 2002 two anti-Indian groups namely Unix Security Guards and World Fantabulous Defacers digitally attacked Indian educational and business sites by carrying out 111 attacks in total when the tension between India and Pakistan increased over Kashmir **[13]** |
| CRISTINA CHIPURICI **[41]** | **Scam** Deceiving people to believe in the things that are not true. Certain fake advertisements or similar kind of slogans are sent to people so as to make them believe in what is being advertised **[41]**. | • Two Nigerians were arrested for duping a woman by promising her a job in America after seeing the profile of the woman which she had uploaded on a well-known job portal. In a mail which she received on December 1, confirming her job in Herald square hotel in New York told her that she would be contacted by the "American embassy", after which she received a mail from the "embassy" asking her to pay the fees and submit her documents in order to go to the states **[12]**.<br><br>• In the year 2012, a scam surfaced which involved Facebook as platform to steal financial information from the users. The attackers impersonated the facebook security account and generated fake messages alerting the users that their accounts would be |
| | | deactivated if they were unable to verify there account. To do so they were directed to a fake page where they were asked to provide their account and credit card details and as a result were defrauded [20]. |
| Neil DuPaul [42] | **Spoofing** Tricking a system to make it trust you when you are actually forging someone else's identity. A person takes up other person's identity in order to penetrate in the system or to shift the blame onto that person [42]. | • An Gujarat ambuja executive pretended to be a girl and blackmailed an Abu Dhabi based NRI for crores [12]. |
| Hon Lau [43] | **Creating malwares** Creating, writing or distributing malicious software for e.g. viruses, worms, Trojans and spywares. These malwares can identify and capture critical information like passwords, username, computer's cache, keystrokes and more [43, 44]. | • A malware that was hosted on Sugarsync cloud services was being distributed by the drive-by-download method. The malicious file was later removed. |
| Ravi Kant [45] | **Spamming** Distributing unwanted e-mails to various email addresses. The content of the mail may be formulated according to the targeted audience and email addresses [45]. | • The CBI registered its first spamming case in 2002 when a complaint was filed by a UK based company regarding thousands of mail that it was receiving from India. The accused was found to be a 16 year old school dropout from Pondichery. |
| Tony Wu Justin Chung James Yamat Jessica Richman | **Wiretapping** Wiretapping is the secret electronic monitoring of internet based communications or telephones, fax etc. It can be accomplished by placing a monitoring device more commonly known as a bug on the wire which we want to monitor or by using built-in mechanisms. For example- packet sniffers are used for acquiring the data that is being transported on any network [32]. | • In 2008 a case was files against the NSA for wiretapping the internet activities of the American citizens. Evidence provided by a former telecommunication technician named Mark Klein who worked for AT&T suggested that AT&T sent the copies of internet traffic to unsaid location in San Francisco which was controlled and monitored by the NSA [19]. |
| Nigel Stanley [47] | **Intellectual property theft** It can be defined as the stealing of any property or material that is copyrighted. The theft of intellectual property- creative ideas, innovation, parts of song or movie, software piracy, trademark violations etc. comes under this category [46, 47]. Around 8% people in India and 7% around the world encountered this attack in the year 2014-2016 [48]. | • Yahoo filed a case in Delhi high court against Akash Arora for using 'yahooindia.com' as a domain name for his website which is very similar to the original website 'yahoo.com' [12]. |

## 2. DIGITAL FORENSICS

Digital Forensics is the branch which deals with the crimes which happen over the computers, where a single computer system constitutes an entire crime scene or in the least it may contain some evidence or information that can be useful in the investigation. However, in technical terms it can be defined as the process of identification, acquisition, preservation, analysis and documentation of any digital evidence [49]. Digital forensics is the process of collecting evidence from any computing device and investigating, analysing and preserving the same to present it as legally admissible evidence in the court of law. The objective of digital forensics is to follow the standardised investigation process while documenting any evidence that is stored digitally which may indicate to the person responsible for the crime. The investigators use various techniques and forensic applications to search hidden folders, retrieve deleted data, decrypt the data or restore damaged files etc. A thorough examination can tell us when any document was created, edited, printed, saved or deleted **[50]**. There are several problems that can be faced by digital forensics examiners like the files that are encrypted take more time, the rapidly changing computer technology, and anti forensics tools can add up to more time and money for the investigating organisation **[51]**. However, as the crime's frequency rises so does its need to get investigated. Therefore, the process which needs to be followed must be thorough and up to its full optimization level in order to solve the case.

## 2.1 PROCESSES INVOLVED IN DIGITAL FORENSICS

The digital forensics process involves the following five levels of investigation **[Harbawi and Varol, 2016]**:

### 2.1.1 IDENTIFICATION OF THE DIGITAL EVIDENCE

This step involves the identification of any digital evidence which might be present at the crime scene. This can involve computers, pen drives, hard disks or any other electronic device that can store digital data. Also, it needs to be taken care of that the processes followed when the computer is found in on or off state are different **[54]**.

### 2.1.2 ACQUISITION OF THE IDENTIFIED EVIDENCE

This step comes after the identification step as after the evidence has been identified it need to be acquired in the most appropriate manner such that the integrity of the data stored in the evidence remains intact. The sub-steps followed during this part of the investigation can be seizing the crime scene, forensically acquiring the data stored in the found devices for further investigation. The two sources of evidence: volatile and non-volatile data have different methods of acquisition. In case of volatile data, the order in which the data is collected is of utmost importance. One suggested order can be network connections, ARP cache, login session, running processes, open files and the contents of RAM. Meanwhile, in case of non-volatile data i.e. from hard-disk the bit stream image can be done using three strategies: using a hardware device such as write blocker where the system is taken offline and the hard drive is removed, using a forensic tool Helix which is used to boot the system or using a live system acquisition which can be done either locally or remotely in case of encrypted systems that cannot be taken offline or are only accessible remotely **[55]**.

### 2.1.3 PRESERVATION OF THE ACQUIRED EVIDENCE

The evidence acquired should be kept in such a way that it remains the same even after the investigation process has been completed as it was first acquired. This is done via a well defines process known as the chain of custody which ensures that the evidence remains protected from the unintended alterations. In order to achieve this, read only copies are made by the practitioners or experts to work upon whereas, the original evidence is kept in a secured location **[]**.

### 2.1.4 EXAMINATION AND ANALYSIS STAGE

This is the most crucial step of any investigative procedure as it is the strongest as well as the most vulnerable part where any minor mistake can lead up to the evidence's ineffectiveness to be presented in the court of law. Examining the evidence involves the step of categorizing the digital evidence and the tools which would be used to analyze that evidence. For instance, a received email can include multiple information regarding the source's email address, metadata as well as the data which can be used to find the ip address of the sender's workstation. One important advice which needs to be taken care of at this stage of analysis involves the difference in the data generated from different devices. For instance, the data and the metadata generated from an image and an email would be different. Evidence's correctness and authenticity to be presented in the court of law totally depends upon the expert's experience and skill **[]**.

### 2.1.5 PRESENTATION OR DOCUMENTATION OF EVIDENCE

This is the last stage of the investigative procedure which includes the process of presenting a report or documentation on what type of evidence was obtained, the description about the experts who worked upon the evidence, the methods followed and the tools used in a specific format. The report can also include the protocols and legal policies followed. It should be presented in the most understandable way stating its findings to be very accurate.

## 2.2 GENERAL PROCESSES INCLUDED IN THE ANALYSIS OF DIGITAL EVIDENCE

Evidence acquired from a crime scene undergoes multiple processes of analysis mentioned as follows **[54, 57]**:

### 2.2.1 VERIFICATION OF OCCURRENCE

The occurrence of the incident which has taken place is verified and the specifics are gathered to figure out the best approach for its investigation. Normally the investigation is done as part of an incident response scenario. The breadth and scope of the incident is determined to assess the case. This preliminary step is important as it helps the investigator in determining the characteristics of the incident and to define the best approach to identify, preserve and collect evidence. **[Rocha et al and Quoinine ,2014,2016]**.

### 2.2.2 DESCRIPTION OF SYSTEM

The data about the specific incident is done by taking notes and describing the system which is to be analysed, where the system is acquired, what is the role of the system in the organization and network. The outlining of the operating system and its general configuration such as disk format, amount of RAM and the location of the evidence is performed.

### 2.2.3 TIME-FRAME ANALYSIS

The information such as when were the files created, modified, changed and accessed in a human readable form is done under the timeline analysis. the timeframe information is also known as MAC time evidence and can be extracted from a variety of tools such as SIFT Workstation. The gathering of data is done using a variety of tools and is extracted from the metadata layer of the file system. it is then parsed and sorted in order to be analysed. The final goal is to generate a snapshot of the activity done in the system including its date, the artefact/media involved, action performed and source.

### 2.2.4 DATA CATEGORIZATION IN MEDIA

The amount of information obtained might be overwhelming at this stage for the investigator and can be reduced by categorizing the files as good or bad. This can be performed by the tools such as hfind from the Sleuth Kit for hash comparisons. The information gives the details about, which files were opened, where did the user browse to, which files were downloaded, which directories were accessed and what programs were executed. Additionally, other things that has to be taken care of is the evidence of account, browser, file downloads, file opening/creation, program execution and usb key usage. The memory analysis is an important analysis step in order to examine rogue processes, network connections, loaded DLLs, evidence of code injection, process paths and many others.

### 2.2.5 INFORMATION FILTERING USING KEYWORD SEARCH

This step is beneficial when one knows what he is looking for. It involves the usage of tools look for the byte signatures of the known files which are also called as magic cookies It is also in this step that one performs string searches using regular expressions. The strings or byte signatures that are searched are the ones that are relevant to the case.

### 2.2.6 RECOVERY OF DATA

This step is done to recover data from the file system. Some of the tools available in the Sleuth Kit can be used to analyse the file system, data layer and metadata layer. The process of analysing the slack space, unallocated space and in-depth file system analysis is part of this step. Carving files based on file headers from the raw images is done to gather evidence. This can be performed using the Foremost tool. The recovered data from the file systems can be analysed using tools such as Sleuth Kit.

### 2.2.7 CAPTURING OF LIVE RAM DUMP

This essential first step is often omitted. Yet, one must begin with acquiring a memory dump before triggering the power switch. Memory dumps contain routinely information such as binary decryption keys for encrypted volumes (TrueCrypt, BitLocker, PGP WDE), recently viewed pictures, loaded registry keys, recent Facebook communications, emails sent and received via Web services such as Yahoo or Gmail, active malware, open remote sessions etc.

### 2.2.8 CREATING DISK IMAGE

Creating a forensic image of the suspect's hard drive is mandatory step in any investigation. A combination of hardware and software is used to acquire forensic disk images while overcoming all possible issues. The hardware is designed to acquire hard drives damaged to the point where the competing imaging products stall. The hardware supports cloning and imaging of a file, enabling one to make up to 3 copies of the source device with a SATA, IDE and USB interface and the receiving device to be a SATA, SSD or USB drive or a file on your computer. You can upload the image onto a remote PC via an Ethernet connection.

### 2.2.9 DISCOVERING AND ANALYZING EVIDENCE

This step might involve the creation of multiple images, out of which most are retrieved. The retrieval of existing and deleted evidence is in auto mode. With hundreds of formats supported, the tools such as Belkasoft Live RAM capturer can quickly extract the most forensically important information. The following types of data can be located or recovered if deleted:

1. Office documents
2. Emails
3. Pictures
4. Videos
5. Mobile application data
6. Web browser histories, cookies, passwords, cache, etc.
7. Chats and instant messenger histories
8. Social network communications
9. System files including jump lists, thumbnails and event logs
10. Encrypted files
11. Registry files
12. SQLite databases
13. PCAP files

### 3. CHAIN OF CUSTODY

Chain of custody is defined as a sequence of events which involves the movement of digital evidence among the authorized experts working upon the evidence in succession. The chain of custody documents the terms related to when, where, who, why, how in the use of evidence at any stage of the investigative process. Its scope involves all the individuals who were part of the process of acquisition, analysis, time records as well as the contextual information i.e. case labelling and the laboratory that processes the evidence **[52]**. Figure 1 shows the relationship between the crime scene, suspect, victim and the chain of custody during any digital forensics investigation.

   The evidence obtained at the crime scene can be of two types: electronic evidence and digital evidence. Electronic evidence refers to any electronic device that can be visually identified and is capable of storing data such as Computer, mobile phone etc whereas Digital evidence refers to the evidence that can be extracted or recovered from electronic evidence (can be any doc file, email attachment, image etc). As depicted in Figure 1 the examination and analysis of the evidence is done by the digital forensic investigator, attorney or lawyer and the judge. And they all after thorough inspection of the evidence present a report which is used for the final verdict of the trial **[53]**.

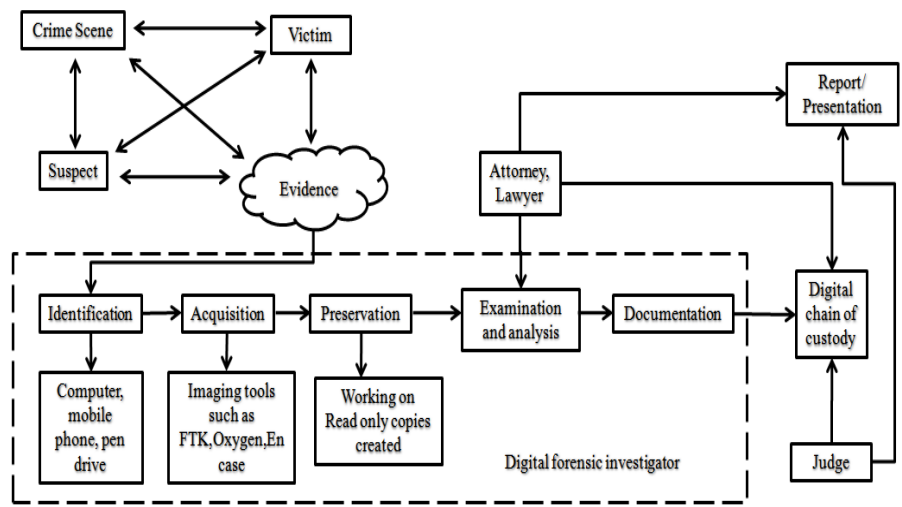

**Figure 1. The relation between the crime scene and the chain of custody**

### 4. COMMONLY USED TOOLS USED DURING DIGITAL FORENSICS ANALYSIS

Digital forensics is one of the many fields that have provided the investigators with multipurpose tools to solve the criminal cases. However, the capability of a single tool to have universal functionality is still rare. The analysis of huge amount of data like that in terms of petabytes in real time and the usage of mobile phones are some of the unforeseen situations that are yet to be adapted by the digital forensics. It would seem most convenient if one tool could be applied to any situation however, it could not be applied to these types of unforeseen advancements. Also, the tools which are multipurpose may provide some tools which are not even required by the investigator and may burden them. The categorization of the forensic tools is done on the basis of the user adopting them. On one hand, law enforcement agencies might require tools which can produce consistent results without compromising the integrity of the evidence whereas, on the other hand Network administrators may require tools that would assist them in the detection of any kind of compromise on their networks. This leads one to conclude two possible truths about digital forensic tools. First that no single tool can act as a universal functionary tool and, second, that all professionals be it investigators, attorneys etc require solutions that provide actionable data consistently. On the basis of these two conclusions, until the offering surfaces forensic toolkits would be providing a vast variety of tools which the investigators can use according to their respective fields **[53]**. The following gives a brief overview about the commonly used tools used by the digital forensic investigators **[88,89]**:

#### 4.1 ENCASE

Encase is a forensic software that is developed by the Guidance Software. It is a 1-mega byte program written in C++ **[58]**. It has a number of products in its suite that includes: Encase cybersecurity, Encase portable, encase forensics and encase eDiscovery. More than 2000 law enforcement agencies use the tool around the world **[58]**. It can be used for variety of purposes including analysis, reporting and acquisition of evidence. The software also has EnScript which is a scripting facility that is used for interacting with the evidence. Encase allows the examiners to identify, search and examine the evidence in order to decide whether further investigation is needed or not, this saves a lot of time and decrease backlog. The tool collect evidences from the file systems, mobile devices and operating systems as well, this ensures that there is no lack of evidence in any case **[59]**.

The investigators place the hard drive of the person in question on the forensic computer; it can be from any operating system like windows, Linux or DOS machine. EnCase makes bit stream mirror image of the drive, which is mounted as read only so that it cannot be altered by anyone. To verify that it is same as original, encase calculates MD5 hash values and redundancy checksums. The tool reconstructs the file structure of the drive and then examines the drive through a Windows GUI. Encase goes deep into the operating system while doing so, in unallocated space or in windows swap files, where deleted files can be stored **[58]**.

### 4.2 BELKASOFT EVIDENCE CENTER

Belkasoft evidence center is a forensic solution that is used for locating, searching, acquiring, examining, sharing and storing the evidence which are stored inside mobile and computer devices. The toolkit extracts the digital evidence from hard drives, memory dumps, blackberry and android backups, IOS, chip-off dumps and more. The tool looks out at hidden locations and for encrypted information for detailed investigation, and carves out damaged or deleted files. It works with different Operating systems like Windows, Linux, Mac, Blackberry and more. It covers more than 700 types of artifacts which include 100 mobile applications, browsers, major documents, pictures and video formats, email clients, social networks, registry files and more **[60]**. It is useful in live RAM analysis and also dig through the cloud data and everything that has been downloaded on that device **[61]**.

### 4.3 CAINE

Caine stands for Computer Aided Investigative environment is a GNU/Linux live distribution that is developed in Italy and is created to act as a Digital forensic project. It provides an environment which is suitable for using the existing forensic tools as well as providing new software modules. It is an open source and available for everyone to install **[62]**. The latest version is Caine 6.0 dark matter which is majorly based on Ubuntu and supports Bios, secure boot etc **[63]**. The main objective of Caine is to provide user friendly graphical interface and tools in an environment that supports digital forensics.

### 4.4 THE SLEUTH KIT AND AUTOPSY

The sleuth kit is a library and collection of UNIX based command line tools that allow an investigator to examine the disk images. The main objective of the sleuth kit to analyze file system data and volume. There are plug-in frameworks that allow a forensic examiner to add on modules to the existing tool and customize the system. The library can be embedded into other forensic tools that can be used directly to find any evidence. The investigators can examine the computer in a non-intrusive manner with the help of the file system tools **[64]**. The toolkit does not depend on the operating system of the device so any data that has been hidden or deleted can be easily viewed **[65]**. Autopsy is the graphical interface that is available with the tools. It has extended features like hash filtering, file system analysis, timeline analysis and keyword searching, adding other modules for increased functionality **[66]**.

### 4.5 FOREMOST

Foremost is an open source digital forensic tool which is designed for Linux based platforms. It is free and was originally developed by Special agents Jesse Kornblum and Kris Kendall of the United States Air Force Office of Special Investigations **[67]**. It is a console program that allows forensic examiners to recover partial or full files from a bit image based on their internal data structure and their headers and footers. This process is also known as data carving. The headers and footers are specified in a user defined configuration file or one can specify the built-in file types by using command line switching **[68]**. Additionally, the configuration files in Foremost allows the investigator to customize the file type that has to be recovered, they can also perform a cursory examination wherein the tool will only search and not extract any file which helps in filtering if there is a lot of media files or drives are involved to determine whether it has to searched thoroughly **[67]**.

Foremost can recover files and images that are created by EnCase, Safeback, the images that are made by dd (Unix utility) or those present directly on the drive **[69]**. The memory portion of media file or image that has to be examined is read by the Foremost tool and is searched for the file header that is specified in the Foremost configuration file. If the header matches then, the tool writes the header and all the data following it to a file until the footer or the size limit of the file specified in the configuration file is reached **[67]**. Foremost can recover files with different extensions namely jpeg, png, gif, exe, bmp, avi, riff, wav, mpg, mov, pdf, wmv, doc, zip, ole, cpp, rar, htm **[70]**.

### 4.6 FORENSIC TOOLKIT

Forensic ToolKit commonly known as FTK is a digital forensic software made by AccessData. It scans and searches the hard drive for any information that the examiner need for example, deleted emails, or scan the disk for text string to use them as possible passwords to crack any encryption. The toolkit has a simple but concise disk imaging program known as the FTP Imager. The FTP Imager saves the image of the hard disk in a file or in segments which can be reconstructed whenever needed **[71]**. The FTK is easy to use and is database driven that is, all the data is stacked in one case database which increases the reusability and reduces the cost and complexity. Any failure that is associated with memory based tools is not encountered in FTK. It is a high speed tool that is designed to give accurate, fast and consistent processing with multithreaded support and distributed processing **[72]**.

The Forensic ToolKit calculates the hash value which is given by the MD5 hash algorithm to confirm the integrity of the achieved data before discarding any file. It gives the result in the form of an image file(s) which can be saved in many formats **[73]**.

### 4.7 MEMGATOR

MemGator is an open source tool, it is easy to use and is portable, cross-platform, concurrent and server tool and self-documented Memento aggregator CLI written in Go. It includes all the features of Memento aggregator and also provides an option to customize different options including which archives are aggregated **[74]**. MemGator is used for memory analysis by extracting the data from memory files automatically and generates a report for the examiner. MemGator combines different memory analysis tools like Volatility Framework, AESKeyFinder, PTFinder and Scalpel file carver into a single program. By using Volatility the user have the options of adding on command line parameters and can override the current program and provide the input of the operating system profile if necessary. Furthermore Scalpel carve and search for usernames and passwords for various social networking platforms like Gmail, Facebook and Yahoo, also will automatically fill the form entry for chrome browser. Additionally it can also take out TrueCrypt encryption keys from the given memory file. The data that can be extracted can be anything related to memory details, passwords, processes, malware detection, network connections and encryption keys. The examination machine should have Python installed in it beforehand **[75]**.

### 4.8 GALLETA

Galleta a word which means "cookie" in Spanish is an open source and free forensic tool that examine and analyse the contents of the cookie files of Internet Explorer that are linked to the browser history and give us the idea of the recently visited websites **[76]**. It was developed as there are many cases where the investigators require the recovery and reconstruction of the Internet Explorer cookie files of the subject. Galleta is a command line tool that is designed to work on various platforms and different operating systems **[77]**. It supports windows, Mac OS, Linux. It analyses the content of the cookie files by converting the data in those files into a format that can be imported to any spreadsheet program for further examination **[78]**.

### 4.9 WIRESHARK

Wireshark is an open source, free and the most widely used packet or network protocol analyser and is a de-facto standard across many educational institutions, government agencies and commercial organisations **[79]**. It was originally known as Ethereal but was renamed in the year 2006. Its development was started in the year 1998 by Gerald Combs **[80]**. Wireshark can run on almost every operating system which includes Microsoft Windows, Linux, Mac OS, Solaris, and BSD. It has another version called TShark which is terminal based **[79]**. Wireshark is used for communication and software protocol development, analysis, network troubleshooting by letting the user see the network at a microscopic level. The data acquired from the network scan can be examined in real time or can also be saved for a later review **[81]**. Wireshark allows the user to have network interface controllers that gives support to promiscuous mode so that all the traffic that is being generated on the network is visible to the user. The packets that are captured are dissected by wireshark that is running on the remote machine at the very same time that they are captured **[80]**.

### 4.10 RIFIUITI

Rifiuti is an open source forensic tool released under the liberal FreeBSD license that is designed to extract information from the recycle bin repository (info2) files of the recycle bin in Microsoft Windows **[76]**. It is written by The word rifiuti literally means "trash" in Italian. It runs on different operating systems like windows, Linux, Mac etc. The recycle bin may have very useful information for the forensic examiner which is why this tool is very useful **[82]**. The tool can recover any recently deleted files from the recycle bin, also it can extract the time when a particular file was deleted, the size of the deleted file, and the original drive number **[83]**. Rifiuti analyses the information in the info2 files and generates the result in a manner that is field delimited so it is easier for the user to export it to any spreadsheet program **[82]**.

### 4.11 NMAP

NMAP which is short for Network Mapper is a free and open source forensic tool which was written by Gordon Lyon. It is used for security auditing and for discovering services and hosts on a network thus giving us a "map" of the network; also many network administrators and systems use it for routine tasks like monitoring service or host uptime and managing service upgrades as well. Although it was initially designed for large networks, it works well against a single host as well. NMAP runs on almost all operating systems like Linux, Windows, Solaris, AmigaOS and Mac and have official binary packages for them, but it started as Linux-only utility **[84]**. NMAP can adapt to any network condition like congestion and latency while doing the scan and is flexible, portable, powerful and well documented. The tool uses specially crafted IP packets and sends it to the targeted host(s) to know what host and services are available on the network, reverse DNS names, what type of firewall is in use, operating system detection, MAC address and much more. Additionally NMAP offers a debugging tool(Ncat), a GUI version(Zenmap) and more **[85]**.

The NMAP generated a result which include "interesting port table" which include information like the service name, port number and protocol etc. The state can be open, closed, filtered or unfiltered. Open means that the target machine has an application that is listening for packets on that port. Closed, means that no application is listening on them. Filtered means that NMAP cannot tell is a port is open or closed because it is being blocked by any network obstacle or firewall. Unfiltered is a state when the NMAP is undetermenistic about the state. The table may also include software version details **[86]**. Nmap hides itself from the source workstation so that the system is not alerted by thinking that it is a malware attack **[76]**.

### 4.12    X-WAYS

X-Ways forensic is an advanced platform that is used by forensic examiners. It is portable and is very flexible to use. The main features of x-ways forensics include disk cloning and imaging, accessing logical memory of running processes, different data recovery techniques, carving of files within other files, identification of lost or deleted partitions, view and edit the binary data structure using templates, ensuring data authenticity by writing protection, hard disk cleansing, analyse the partition table without altering the original content, detection and much more **[87]**.

### 5.  CHALLENGES OF DIGITAL FORENSICS

There is no simple answer to the question about the challenges faced by the digital forensics today. As a field digital forensic is encountering immense array of challenges out of which not even a single one is candidly answerable. Researchers and experts from around the globe coincide together to foresee the elucidation of the major challenging facet of the field **[90]**.
In a survey conducted by the Forensic Focus in September 2015, around five hundred people were asked to voice their opinion on the biggest challenges faced today by the digital forensic investigators. This question stimulated overabundance of answers. The research done by the experts from University College Dublin's School of Computer science have questioned the same **[91]**.The result of this survey revealed that the two most concerning factors were encryption(21%) and cloud forensics(23%) followed by triage(11%), lack of resources and training, the increased volume of data, the elevation in the number of digital crimes etc **[90]**.

#### 5.1 Device proliferation

Mobile and IoT equipments use array of operating systems, communication specifications and file arrangements, which in turn increases the complexity of investigations. In few cases there may be an absence of constant storage wholly, postulating pricey RAM forensics. Furthermore inspecting many devices also adds up to the correlation and consistency problem **[91]**.

#### 5.2 Lack of training and resources

Lack of training and resources has also been listed as a major challenge. The already strained investigators check all the evidence from multiple devices manually, which becomes a matter is concern. Only a basic training is given to the investigators about using a particular forensic tool without giving them any deeper knowledge of its working and techniques **[90]**. Around 44% people in India thought that the local law enforcement agencies lack the required knowledge and skills in handling the cases related to cyber crimes, malware hacking incidents etc **[6]**.

### 5.3 Cloud forensics

Unlike the traditional forensic scheme where the data is stored on a single machine, the data on the cloud is stored on different nodes. This leads to data being in different jurisdiction and the investigators facing local laws and codifications for the collection of any evidence. This may possibly increase the amount of time, expenses and predicament correlated with the forensic investigation. The matter that an individual file can be broken down into multiple blocks of data that are stored in isolated nodes summates the complexity. The research conducted by the UCD also agrees on the challenges regarding the cloud-based data storage.
A law student,Janice Rafraf studying at Teesside University, conveyed through a presentation at TDFCON in 2015, not only the most evident obstacles like accessing the data stored in unfamiliar surroundings but also the legal struggle related to it. The data may not be legally accessible for it is stored in many distinct physical locations globally.
The diverse nature of data storage makes the covering of tracks for criminals effortless with all the tools available for keeping the person anonymous. The adoption of IP obscurity tools along with the user friendly aspect of the cloud systems can make it impossible in tracking down a criminal **[92, 93]**.

### 5.4 Encryption

When asked about the impact of encrypted devices on investigations Yuri Gubanov, CEO of Belkasoft said that the challenges differ between devices. For example in windows a full disk encryption can be invaded by seizing a memory dump in order to get the binary decryption key .For androids it hinges on the version of android and the creator of the device, they can be decrypted or dumped even if it is pass code protected. The enforcement of security is laudable in the Apple tablets and phones; it is using Secure Enclave in 64-bit hardware for the purpose. Other aisle like cloud is used in case of data acquisition in such device. The real challenge is to actually locate the encrypted data; it is difficult to distinguish between the compressed files and the data which is encrypted. The experts are applying different procedures (workarounds) to triumph encryption. For example, if one knows the precise password of the Microsoft account, they can unlock a BitLocker volume or by obtaining a binary decryption key from a memory dump. One can easily identify the weakness of any android Smart phone in order to surpass its protection or can get more than enough data from the Google account of the user. The information can easily be extracted from the cloud in case of Apple smart phones where it is backed up by default instead of breaking into the device. Once all the paramount data is acquired it necessary to analyze what is useful for the investigation and discard the rest. The estimated increase in the case volume would add to the already elevating backlog problem especially if the amount of evidence originating from cloud and Internet of Things keeps on escalating. Forensic professionals frequently find themselves despondently mangling to keep up to the ever growing development of new technologies. With the surge of the society stockpiling heap of data online, it can serve as an opening for the field of digital forensic to use this enthusiasm as a leverage and help to devise compelling solutions. Collusion between scholastic establishments, law enforcement agencies and business organizations are unarguably the best way to do so **[90]**.

### 5.5 Legal Issues

There is a stress on the privacy and protection of data of all citizens, which has been evolving recently. Everyone is the owner of their data and devices. This factor adds to the complexities faced by the forensic examiners for gathering the required information. For example any information that may be of use is contained within the suspect's device cannot be achieved because of the laws that change with the change in the geographical areas and searching for information from that device may be considered violating the rights of the person and violation of rules. The fact that now day the organizations are allowing their employees to do the official work from their personal devices also adds up to the stack of problems. If there is any malicious activity performed by that device will be difficult to trace back like if someone downloads an attachment from any email that can result in theft of confidential information **[94]**.

### 6. PROPOSED SOLUTIONS AND FUTURE SCOPE

The increase in the number of cybercrimes has engendered us to focus upon the improvement of the current legislation and effectiveness of the current forensic tools in the investigation of cybercrimes. Even today, the first responders tend to ignore the importance of the digital evidences found on the crime scene. In most of the cases, the judgement is merely made on the basis of circumstantial evidences which may be proved wrong after the inclusion of the digital evidence as part of the investigation. For this, we suggest that the law enforcement agencies organise certain basic training programs for their first responders so that they do not neglect the digital evidence and acquire it in a forensically correct manner that is by not unplugging the system directly and create a necessary forensic image of the volatile memory also known as RAM forensics. In addition to the training programs and the suggested acquisition measures, the process of image forgery detection needs to be improved in investigation of cybercrimes in India. So that instead of judging the image's authenticity on the basis of the person's skin tone it can be validated on the basis of its embedded signature or other pixelating factors.

### 7. CONCLUSION

In this paper, we briefed about the status and the various types of cybercrimes happening worldwide. We then, discussed about the importance of digital evidence and digital forensics processes involved in the cybercrime investigation. Thereafter, we mentioned about the chain of custody process adopted in the digital forensic investigation and about the various actors who play key roles in it. The process was then followed by a detailed explanation about various digital forensics tools used in the investigations. The detailed study of these tools also highlighted the challenges associated with the field of digital forensics. Towards the end of the paper, possible solutions were suggested as countermeasures for the problem under consideration.